\documentclass[reprint,aps,nofootinbib,superscriptaddress,preprintnumbers]{revtex4-2}
\usepackage{comment}
\usepackage[bottom]{footmisc}
\usepackage{graphicx} % Required for inserting images
\usepackage{braket}
\usepackage{amsmath,amssymb,amsfonts}

\usepackage{dcolumn}                % Align table columns on decimal point
\usepackage{bm}                     % Bold math
\usepackage{ulem}

\usepackage{mathtools}
\usepackage[dvipsnames]{xcolor}

\newcommand\blfootnote[1]{%
  \begingroup
  \renewcommand\thefootnote{}\footnote{#1}%
  \addtocounter{footnote}{-1}%
  \endgroup
}

\begin{document}

\preprint{CALT-TH 2025-021}

\title{Observer Time from Causality in Perturbative Quantum Gravity}

\author{Allic Sivaramakrishnan}
\affiliation{Walter Burke Institute for Theoretical Physics, California Institute of Technology, Pasadena, CA}

\blfootnote{allic@caltech.edu}

\begin{abstract}
It is unclear whether an observable notion of time exists in quantum gravity even in principle because spacetime itself fluctuates. We propose a form of observable time in perturbative quantum gravity. First, we define an elapsed proper time in curved space using intersections of worldlines and future light cones. Here, time arises from causality via the dependence on light cones. We then propose that performing the gravitational path integral describes quantum gravity corrections to this notion of proper time at all orders in $G_N$. Using this prescription, we compute the leading quantum gravity corrections to two-point correlation functions of elapsed time. We find that the commutators can be nonzero, showing this notion of time is a quantum operator in quantum gravity.

\end{abstract}

\maketitle
%\tableofcontents

\section{Introduction}

What an observer would see in the presence of quantum gravity effects, at least in principle, is operationally defined by what they can measure. However, defining measurements can be subtle because even in the vacuum, spacetime fluctuations may invalidate the underlying geometry. Also, measurements of a quantum system generally influence the system, physical observers gravitate, and observables can be observer-dependent in classical and quantum gravity (see \cite{DeVuyst:2024pop, DeVuyst:2024uvd, Chandrasekaran:2022cip} for recent work).

Realistic observables can be defined in relation to a measurement apparatus or observer, whose dynamics are included explicitly. This relational definition makes the coordinate-invariance of observables manifest. Common models of observers are ruler and clock fields, and also the worldlines of point particles. The observer degrees of freedom can be quantized via canonical quantization or path-integral methods. See \cite{Giddings:2025xym} for a modern overview.

Relational observables can be defined in the low-energy effective field theory (EFT) of quantum gravity. Here, small perturbations around a background spacetime are quantized according to the well-established rules of perturbative quantum field theory \cite{Donoghue:1995cz,Burgess:2003jk}. Predictions can be made for quantum gravity signatures in highly-sensitive laser interferometers \cite{Kanno:2020usf,Parikh:2020fhy,Parikh:2020kfh,Parikh:2020nrd,Zurek:2020ukz, Verlinde:2019xfb, Carney:2024wnp}, for instance the LIGO \cite{aasi2015advanced} and GQuEST \cite{Vermeulen:2024vgl} experiments. Applying EFT methods to realistic observables can expose how geometric quantities like time, location, and distance experience the onset of quantum gravity effects.

In this work, we study the proper time $\Delta \tau$ elapsed between two signals that intersect an observer's worldline. In the classical regime, the experimental observable in laser interferometers can be understood as such a time measurement \cite{Misner1973, ligo-note, Maggiore:2007ulw}. This definition of $\Delta \tau$ has been extended to linearized quantum gravity \cite{Carney:2024wnp}. 

We propose a definition of elapsed time $\Delta \tau$ valid to all perturbative orders in the EFT of quantum gravity. The principle underlying our construction is that causality furnishes a notion of time. In Section \ref{protocol}, we define proper-time correlators in perturbative quantum gravity. On a fixed background $g_{\mu\nu}$, the intersections between worldlines and the future light cones of points along those worldlines define elapsed proper times $\Delta \tau$ of worldline segments in a diffeomorphism-invariant way. This radar-like light clock can be implemented by observers carrying lightbulbs. Then, expanding around a background metric $g^{(0)}_{\mu\nu}$ via $g_{\mu\nu} = g_{\mu\nu}^{(0)}+h_{\mu\nu}$ and performing the gravitational path integral over the perturbations $h_{\mu \nu}$ defines correlators of $\Delta \tau$ to all perturbative orders in $G_N$. In Section \ref{propertimecorrelators}, we compute the leading quantum gravity corrections to $\braket{0|\Delta \tau(t) \Delta \tau(0)|0}$, the unequal-time vacuum two-point correlation function. In flat space, $g^{(0)}_{\mu\nu} = \eta_{\mu\nu}$, the leading correction to $\braket{0|\Delta \tau(t) \Delta \tau(0)|0}$ is $\mathcal{O}(h^2)$, or $\mathcal{O}(G_N)$. To obtain all the non-zero contributions to this correlator, we compute $\Delta \tau$ to $\mathcal{O}(h^2)$. We then compute the commutator in closed form and find that $\braket{0|[\Delta \tau(t), \Delta \tau(0)]|0} \neq  0$ for certain $t$, indicating that $\Delta \tau$ is a quantum operator. Similar results hold for a clock synchronization protocol. Our results reproduce and generalize those of \cite{Carney:2024wnp}, which analysed $\Delta \tau$ through $\mathcal{O}(h)$.

\section{Proper time as a quantum operator}
\label{protocol}

\subsection{A lightbulb clock}
\label{lightbulbclock}

We first construct a classical clock on a curved background. We model observers as point masses following worldlines $x(\tau)$, which we parametrize by proper time $\tau$. We suppose the worldlines originate with initial conditions for $x, \frac{d x}{d \tau}$ at some location, for example the far past, where gravity is non-dynamical and the diffeomorphisms we consider do not act. This is sometimes called a platform \cite{Donnelly:2015hta}. Operations along the worldline occur at times $\tau$ elapsed from the platform, which is a diffeomorphism-invariant designation. See \cite{Donnelly:2015hta, Donnelly:2016rvo, Goeller:2022rsx} for details.

Consider two inertial worldlines, $x_1, x_2,$ in a sufficiently well-behaved spacetime, for instance with no curvature singularities or trapped surfaces. A lightbulb is turned on at time $\tau_1$ along worldline $x_1$. Its future light cone, the set of points $L^+[x_1(\tau_1)]$, intersects the worldline $x_2$ exactly once: $L^+[x_1(\tau_1)] \cap x_2  = x_2(\tau_2)$ for some $\tau_2$. The intersection condition defines $\tau_2$ as a diffeomorphism-invariant function of $\tau_1$. $\tau_2$ is the first possible moment at which the observer $x_2$ can see light from the lightbulb. That such a moment exists is due to causality in that signals cannot propagate faster than the speed of light.

Similarly, a lightbulb turned on at $x_2(\tau_2)$ has a future light cone $L^+[x_2(\tau_2)]$ that intersects $x_1$ exactly once. That is, $L^+[x_2(\tau_2)] \cap x_1  = x_1(\tau_3) $ for exactly one $\tau_3$, which is the earliest possible time the second lightbulb's signal can reach $x_1$. 

\begin{figure}[ht]
    \centering
    \includegraphics[width=0.30\textwidth]{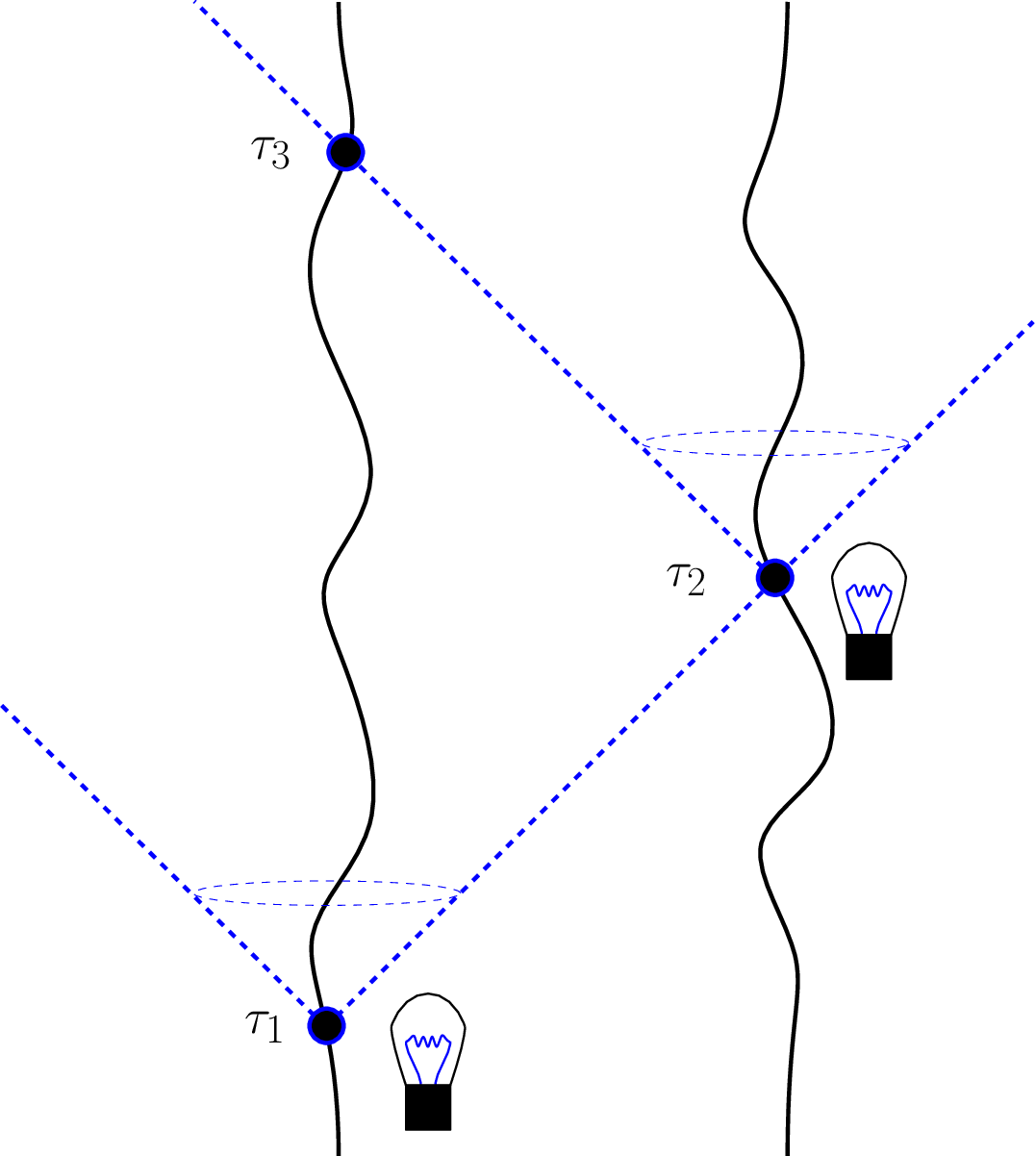}
    \caption{One ``tick'' of a lightbulb clock built from inertial observers (solid lines) and future light cones (dotted lines). One observer turns a lightbulb on at their proper time $\tau_1$. The signal reaches the second observer at their time $\tau_2$, at which point the second observer immediately turns on their own lightbulb. This new signal reaches the first observer at their time $\tau_3$. This procedure defines the elapsed proper time $\tau_3-\tau_1$ in a diffeomorphism-invariant way in curved space. Light cones are drawn as straight lines for clarity.}
    \label{lightbulb_clock}
\end{figure}

The process we have described defines $\tau_3-\tau_1$ as one ``tick'' of a lightbulb clock. See figure \ref{lightbulb_clock}. This clock is standardized at the time of initialization, but not afterwards. Here, $\tau_3$ is a function of $\tau_1$, the initial data of both worldlines, and the metric.

We can compute $\tau_3-\tau_1$ as follows. The intersection times of the light cones and worldlines can be computed by identifying the null geodesic $x_p$ that connects the worldlines $x_1, x_2$. Denoting the outbound and return null geodesics by the piecewise-defined $x_p$, we solve the equations of motion for this system,
\begin{align}
&\frac{d^2x_i^\mu}{d\lambda_i^2}  + \Gamma_{\rho \sigma}^\mu \frac{dx_i^\rho}{d\lambda_i}\frac{dx_i^\sigma}{d\lambda_i} = 0,
\nonumber
\\
&\frac{d^2x_p^\mu}{d\lambda_p^2}  + \Gamma_{\rho \sigma}^\mu \frac{dx_p^\rho}{d\lambda_p}\frac{dx_p^\sigma}{d\lambda_p} = 0, ~~~~~~ \left(\frac{dx_p^\mu}{d \lambda_p}\right)^2 = 0.
\label{WorldlineEOMs}
\end{align}
where $x_i = x_i(\lambda_i)$ and $x_p = x_p(\lambda_p)$. We use the boundary conditions
\begin{equation}
x_p(\lambda_p)=x_1(\tau_1),
~~~x_p(\lambda_p')=x_2(\tau_2),
~~~
x_p(\lambda_p'')
=x_1(\tau_3),
\end{equation}
for some values $\lambda_p < \lambda_p' < \lambda_p''$ of affine parameter. One then solves for $\tau_2, \tau_3$ in terms of $\tau_1$. 

In this section, we have described a variant of the well-known light clock and radar times adapted to curved space. The standard light clock uses light rays and so in curved space is valid only approximately when the distances between $x_1, x_2$ are sufficiently small \cite{Misner1973}, although see \cite{Perlick1987,Perlick:2007np}. The phase shift in laser interferometers can be understood as $\tau_3-\tau_1$ for the linearized metric perturbations that describe gravitational waves \cite{Misner1973, ligo-note, maggiore2008gravitational}, but is less well studied at higher orders. The lightbulb clock discussed here appears valid for a wide range of curved backgrounds, does not require reflection, and can be implemented by realistic observers. This clock is an idealized geometric construction but more realistically is inferred from detector correlators (see Appendix \ref{detectors}).

\subsection{Path-integral definition of proper-time correlators}
\label{pathintegral}

The lightbulb clock is valid whether or not the background geometry satisfies Einstein's equations. To include quantum gravity corrections to elapsed time, we propose that we expand about a fixed background $g^{(0)}_{\mu\nu}$ as $g_{\mu\nu} = g^{(0)}_{\mu\nu} + h_{\mu\nu}$, and then perform the path integral over the perturbatively small fluctuations $h_{\mu\nu}$. Crucially, we assume that when $h_{\mu\nu}$ is small, it does not violate the lightbulb clock conditions at any perturbative order. Standard EFT methods then compute perturbative quantum gravity corrections to elapsed time. 

As both $\tau_{2}, \tau_{3}$ are functions of $h_{\mu\nu}$ and $\tau_1$, their one-point functions are
\begin{equation}
\braket{\tau_{3}(\tau_1)} \equiv \int \mathcal{D}[h] \tau_{3}(\tau_1,h) e^{-iS[h]},
\label{timeonepoint}
\end{equation}
and similarly for $\braket{\tau_2(\tau_1)}$, with 
\begin{equation}
S[h] = \frac{1}{16 \pi G_N} \int d^4x \sqrt{-g} R+ \sum_i m_i\int d\tau_i,
\label{action}
\end{equation}
where the four-dimensional action above is evaluated at $g_{\mu\nu} = \eta_{\mu\nu} + h_{\mu\nu}$ and expanded in small, dimensionless $h_{\mu\nu}$, or equivalently in $G_N$. One may also rescale $h_{\mu\nu} \rightarrow \sqrt{32 \pi G_N} h_{\mu\nu}$ so that $h$ has dimensions of inverse length. The gauge-fixing factor in the path integral is left implicit for brevity. In \eqref{timeonepoint}, $\tau_1$ is a parameter rather than determined by some dynamical process, or by $h$. The expectation value of elapsed time is $\braket{\tau_3(\tau_1)-\tau_1} = \braket{\tau_3(\tau_1)}-\tau_1$.
 The two-point function is
\begin{equation}
\braket{\tau_3(\tau_1) \tau_3(\tau_1')}
=
\int \mathcal{D}[h] \tau_3(\tau_1,h)\tau_3(\tau_1',h) e^{-i S[h]},
\end{equation}
and similarly for $\tau_2$. This determines the two-point function of elapsed time, $\braket{(\tau_3(\tau_1)-\tau_1)(\tau_3(\tau_1')-\tau_1')}$.

The times $\tau_2, \tau_3$ have become quantum operators.\footnote{Fluctuations in $\tau_3$ imply the same for the causal diamond defined by $x_1(\tau_1), x_1(\tau_3)$. The associated modular Hamiltonian is central to a formal notion of time via modular theory \cite{Takesaki:1970aki}. } Imposing the equations of motion for the worldlines and performing the path integral over $h_{\mu\nu}$ was a procedure studied in \cite{Donnelly:2015hta}. The main novel feature of our work is to define proper-time correlators in this setup.

\section{Proper-time correlators}
\label{propertimecorrelators}

We now compute the two-point correlators of proper time defined in Section \ref{protocol} to lowest order in $G_N$ and study their properties. Each factor of $h$ will contribute a factor of $\sqrt{G_N}$ and we will find that the leading correction to the two-point function is $\mathcal{O}(h^2)$, or $\mathcal{O}(G_N)$. We will therefore need to compute elapsed time, specifically $\tau_3$, through $\mathcal{O}(h^2)$. We consider an arbitrary perturbation $h_{\mu\nu}$ about flat space, $g_{\mu\nu} = \eta_{\mu\nu}+h_{\mu\nu}$.

First, we solve the equations of motion in \eqref{WorldlineEOMs} perturbatively in $h$. We expand the timelike worldlines as $x_i = x_i^{(0)}+x_i^{(1)}+x_i^{(2)} + \cdots,$ where $x_i^{(n)}$ is $\mathcal{O}(h^n)$. Using coordinates $x^\mu = (t,x,y,z)$, we choose
\begin{align}
&x_1^{(0)}(t) = (t,0,0,0), ~~~ x_2^{(0)}(t) = (t,0,0,L),
\\
&
x^{(0)}_p (t_1, \lambda) = (\lambda+t_1, 0,0,\theta(L-\lambda) \lambda + \theta(\lambda-L)(2L-\lambda)).
\nonumber
\end{align}
Here $L$ is specified via initial conditions and so is diffeomorphism invariant, and $\lambda \in [0,2L]$. We have chosen coordinates such that $t_1$ coincides with the proper time $\tau_1$. We work in synchronous gauge, $h^{00}=h^{i0} = 0$, where Latin indices run over $x,y,z$. We can therefore use $\lambda=t$ as the affine parameter. In synchronous gauge, $x_1^{(n)} = x_2^{(n)} = 0$ for all $n \geq 1$, which can be proven by perturbatively solving $\ddot{x}_i^\mu+ \Gamma^\mu_{\rho\sigma} \dot{x}^\rho_i \dot{x}^\sigma_i = 0$. 

Next, we compute $x_p$. The exact metric is
\begin{equation}
ds^2 = -dt^2 +(\delta_{ij}+ h_{ij})dx^i dx^j.
\end{equation}
The trajectory of the photon is $ds^2=0$, or $1=(\delta_{ij}+ h_{ij})v^i v^j$ with $v^i = dx^i_p/dt$, and is expanded as $v_i = v_i^{(0)} + v_i^{(1)}+v_i^{(2)} + \cdots$. We then solve $ds^2=0$ perturbatively to determine $v^{(1)}, v^{(2)}$. See Appendix \ref{propertime} for the results.

With the solutions to the equations of motion in hand, we can now compute $\tau_2,~\tau_3$ perturbatively in $h$. We expand $\tau_i = \tau^{(0)}_i+\tau^{(1)}_i+\tau^{(2)}_i$. In synchronous gauge, $\tau_2=t_2, \tau_3=t_3$. We now must solve for the times $t_2, t_3$ at which $x_p$ intersects the $x_i$. The defining conditions are
\begin{align}
L &= \int_{t_1}^{t_2} v_z ~d t 
=  \int_{t_1}^{t_2}  \left(1 - \frac{1}{2} h_{zz} + v_z^{(2)}\right)dt,
\nonumber
\\
-L &= \int_{t_2}^{t_3} v_z~dt 
=\int_{t_2}^{t_3} \left(-1 + \frac{1}{2} h_{zz}+v_z^{(2)}\right)dt,
\label{definingconditions}
\end{align}
where $v_z^{(0)} v_z^{(1)} = -\frac{1}{2}h_{zz}$. Subtracting these equations,
\begin{equation}
t_3-t_1 = 2L + \int_{t_1}^{t_3} \left(\frac{1}{2} h_{zz}- v_z^{(0)} v^{(2)}_z \right) dt,
\end{equation}
where we introduced $v_z^{(0)}$ to encode a sign. We then find
\begin{align}
t_3^{(0)} &= 2L +t_1,
\\
t_3^{(1)} &= \frac{1}{2}\int_{t_1}^{t_1+2L} h_{zz}(x_p^{(0)}(t_1,t))dt,
\nonumber
\\
t_{3}^{(2)} &= \frac{1}{2}t_3^{(1)} h_{zz}(x_p^{(0)}(t_1,t_1+2L))-\int_{t_1}^{t_1+2L} v_z^{(0)}v^{(2)}_z dt,
\label{propertimeperturbativecorrections}
\end{align}
where $v_z^{(2)}$ is given in Appendix \ref{propertime}. The $t_3^{(1)}$ given above agrees with \cite{Lee:2024oxo}, which computed it in an arbitrary gauge.

Computing $\tau_3-\tau_1$ up to $\mathcal{O}(h^2)$ enables us to determine elapsed time correlation functions at $\mathcal{O}(h^2)$. Denoting the function $\tau_3(\tau_1)-\tau_1 \equiv \tau(t_1)$ for simplicity,
\begin{align}
&\braket{\tau(t)\tau(0)} = 4L^2 + \braket{\tau(t) \tau(0)}_c
\nonumber
\\
&
~~~~+2L( \braket{\tau^{(1)}(t)}+\braket{\tau^{(2)}(t)}+\braket{\tau^{(1)}(0)}+\braket{\tau^{(2)}(0)}),
\label{propertimetwopointfunction}
\end{align}
where the connected correlator, or the covariance, is defined as $\braket{\tau \tau'}_c = \braket{(\tau-\braket{\tau})(\tau'-\braket{\tau'})} = \braket{\tau \tau'}-\braket{\tau}\braket{\tau'}$. This can be understood as capturing measurements made by two different lightbulb clocks. As $\tau$ consists of non-local integrals of $h$, the EFT prescription in Section \ref{pathintegral} allows us to analyse each term in \eqref{propertimetwopointfunction} through $\mathcal{O}(h^2)$.

From the form of $v_z^{(2)}$ in Appendix \ref{propertime}, we see the $\braket{\tau^{(2)}}$ 
contributions are graviton corrections to the null ray. These are various self-corrections and do not couple the two lightbulb clocks. As expected, these diagrams don't contribute to the connected proper-time correlator. 

The $\braket{\tau^{(1)}}$ terms through $\mathcal{O}(h^2)$ are
\begin{align}
\braket{\tau^{(1)}(t)} &= \frac{1}{2} \int dt' \braket{h_{zz}(x_p^{(0)}(t,t'))} 
\nonumber
\\
&= \frac{1}{2} \int dt' \int d^4 y  P_{zz\mu\nu}(x_p^{(0)}(t,t'),y)T^{\mu \nu}(y),
\label{response}
\end{align}
where the $\mathcal{O}(h)$ term was proportional to the vacuum one-point function, $\braket{h_{zz}} =0$, and $T^{\mu\nu}(y) = \frac{1}{2}\sum_i m_i \int d\lambda_i \dot{x}_i^\mu \dot{x}_i^\nu \delta^{(4)}(y-x_i)$ is the worldline stress tensor at leading order in $h$. The $P_{\mu\nu\rho\sigma}(x,y) \sim \mathcal{O}(h^2)$ is the synchronous-gauge Feynman propagator. The terms in $\braket{\tau^{(1)}}$ describe how the gravitational force from the worldlines of either lightbulb clock affects a single proper time measurement. These potential terms are proportional to $m_i$, and are precisely the proper-time response functions in \cite{Sivaramakrishnan:2024ydy}. The diagrams are connected in that they couple lightbulb clocks, but the coupling does not induce any correlation between the time measurements.

\begin{figure}[ht]
    \centering
    \includegraphics[width=0.35 \textwidth]{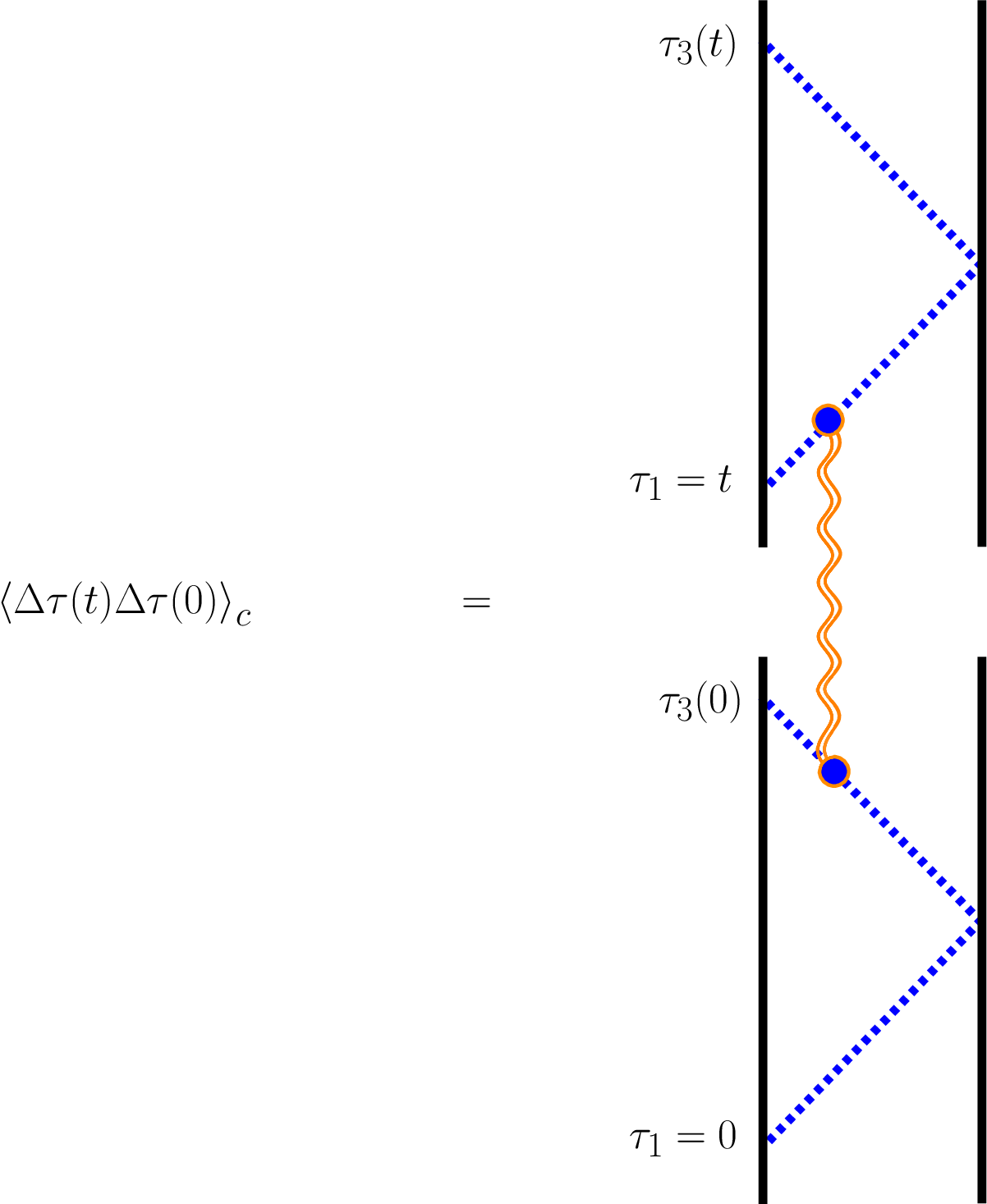}
    \caption{The diagram that computes the connected two-point function of elapsed time $\braket{\Delta \tau(t) \Delta \tau(0)}_c$ at leading order, $\mathcal{O}(G_N)$. Here, $\Delta \tau(\tau_1) = \tau_3(\tau_1)-\tau_1$. The graviton two-point function (double wavy line) is integrated over the unperturbed null rays (dotted lines) between the worldlines (solid lines). In synchronous gauge, the worldline locations are unchanged by the metric fluctuations. We have depicted two distinct experimental apparatuses for conceptual clarity.}
    \label{2pt_function}
\end{figure}

The remaining term is $\braket{\tau(t) \tau(0)}_c = \braket{\tau^{(1)}(t) \tau^{(1)}(0)}_c$, which through $\mathcal{O}(h^2)$ is 
\begin{equation}
\braket{\tau(t) \tau(0)}_c =  \frac{1}{4}\int dt' \int dt'' \braket{ h_{zz}(x_p(t,t')) 
 h_{zz}(x_p(0,t'')) 
}.
\label{tautaubackgroundfluctuation}
\end{equation}
Based on our analysis of the terms in \eqref{propertimetwopointfunction}, we conclude that $\braket{\tau(t)\tau(0)}_c$ describes the effect of correlated metric fluctuations. See figure \ref{2pt_function}.

We consider the Wightman function defined as the $\epsilon \rightarrow 0^+$ limit of $\braket{\tau(t-i\epsilon)\tau(0)}$. Because free Wightman functions are sums over on-shell plane waves, we use the residual gauge freedom in synchronous gauge to satisfy the transverse-traceless gauge conditions, which means \eqref{tautaubackgroundfluctuation} is precisely the proper-time correlator defined in \cite{Carney:2024wnp}. 

The correlator \eqref{tautaubackgroundfluctuation} is the leading order quantum gravity correction to the elapsed-time observable in a toy model of a laser interferometer \cite{Carney:2024wnp}. Here, a null ray $x_p$ starting from $x_1(\tau_1)$ is aimed to reach the mirror, $x_2$, then reflects and returns to $x_1$. This requires foreknowledge of the metric, and metric perturbations generically cause $x_p$ to miss $x_2$, which renders the round-trip time ill-defined. The procedure in Section \ref{lightbulbclock} avoids these issues. As the signal in \cite{Vermeulen:2024vgl} is the modulation of a carrier frequency, we expect that the connected correlator models the interferometer observable.

As reviewed in Appendix \ref{propertimeoperator}, $\braket{\tau(t) \tau(0)}_c$ is \cite{Carney:2024wnp} 
\begin{align}
&\braket{\tau(t) \tau(0)}_c
=\frac{16  G_N}{\pi} \lim_{\epsilon \rightarrow 0^+} 
\int_0^\infty \frac{dk}{k} e^{-i k (t-i\epsilon)} 
\nonumber
\\
&\times \Big[ \frac{3 - \cos(2 k L)}{6}  - \frac{3 + \cos(2 k L)}{2 k^2 L^2} + \frac{\sin(2 k L)}{k^3 L^3} \Big].
\end{align}
One signature feature of a quantum operator is its failure to commute. The full commutator is $\braket{[\tau(t),\tau(0)]} = 2i\, \text{Im}\braket{\tau(t)\tau(0)}_c$. Evaluating the right-hand side and using the dimensionless coordinate $u = t/2L$, we find that $\braket{[\tau(t),\tau(0)]} = 0$ for $u > 1$. For $0<u<1$,
\begin{equation}
\braket{[\tau(t),\tau(0)]} = -i \frac{8 }{3}G_N(3 - 12 u + 18 u^2-8 u^3).
\label{tautaucommutator}
\end{equation}
We conclude that proper time is a genuine quantum operator. $\braket{[h(x_1),h(x_2)]} =0$ away from the light cone (e.g. see \cite{Ooguri:1985nv}), explaining why $\braket{[\tau(t),\tau(0)]} = 0$ for $u>1$. That $\braket{[\tau(t),\tau(0)]} \neq 0$ for a range of $t$ confirms that $\tau$ is intrinsically nonlocal. As a consistency check, note that the only two diffeomorphism-invariant quantities in the setup are $L$ and proper time $\tau_1$, or $t$ in this gauge, and indeed $\braket{[\tau(t),\tau(0)]}$ is $G_N$ multiplying a function of the gauge-invariant, dimensionless quantity $u$.

Initializing the worldlines somewhere that gravity was non-dynamical rendered the setup well-defined. In practice, the two worldlines are usually defined by measurements during which gravity is dynamical. As a simple example, a lightbulb turned on at the spacetime point $x_2(\tau_2)$ allows the time $\tau_2$ to be synchronized with $\tau_3$. These are related by $t_3 = t_2+L +\frac{1}{2}\int_{t_2}^{t_2+L} h_{zz} dt+\mathcal{O}(h^2)$. In this case, using transverse-traceless gauge we have
\begin{align}
\braket{\tau_3(t) \tau_3(0)}_c
=&\frac{4 G_N}{\pi } \lim_{\epsilon \rightarrow 0^+} 
\int_0^\infty \frac{dk}{k} e^{-i k (t-i\epsilon)} 
\nonumber
\\
&\times \Big[\frac{4}{3}-\frac{2}{k^2 L^2} +\frac{ \sin(2 k L)}{k^3L^3}\Big],
\end{align}
which is implicit in \cite{Carney:2024wnp}. For $0<u<1$,
\begin{equation}
\braket{[\tau_3(t),\tau_3(0)]}= i\frac{16 G_N}{3}(u-1)^3,
\end{equation}
and $\braket{[\tau_3(t),\tau_3(0)]}=0$ for $u>1$. Even if an observer can signal at a definite time, we see that because spacetime fluctuates during the entire process, quantum gravity endows the signal's arrival time with quantum properties. Evidently, we cannot always synchronize clocks in perturbative quantum gravity even in this simple model.

To determine the full Wightman function, we also need the anticommutator, 
\begin{align}
    \braket{ \left \{ \tau(t),\tau(0) \right \}}
=2\,\text{Re}\braket{\tau(t)\tau(0)}_c&+4L\big( \braket{\tau^{(1)}(t)}+\braket{\tau^{(2)}(t)}
    \nonumber
\\    
   &+\braket{\tau^{(1)}(0)}+\braket{\tau^{(2)}(0)}\big).
\end{align}
We provide the first term in Appendix \ref{anticommutator}. Noting that unitarity and on-shell methods apply to correlators in curved space \cite{Meltzer:2020qbr, Cheung:2022pdk}, we leave further study of the anticommutator to future work. 

\section{Future directions}

We have proposed a definition of proper-time correlators in perturbative quantum gravity and a general approach to constructing other geometric observables. A number of avenues are open. 

The ideas presented here, including extensions to loops, excited states, and AdS/CFT, may be used to investigate any of the experimental signatures studied in \cite{Kanno:2020usf, Parikh:2020fhy,Parikh:2020kfh,Parikh:2020nrd,Zurek:2020ukz, Verlinde:2019xfb, Aalsma:2025bcg} that arise within EFT. Including the worldline path integrals and detector operators may clarify to what extent realistic measurements intrinsically gravitate. Charged particles coupled to vacuum fluctuations of a gauge field may experience analogous correlations in phase shifts and time delays.

We used a standard, experimentally-relevant worldline model and an associated path-integral description. The success of high-loop computations of scattering amplitudes indicates this path-integral approach is well-suited to studying perturbative corrections, including a rigorous treatment of ultraviolet and infrared divergences. It may be worth exploring whether the approach we used agrees even perturbatively with other definitions of subregions \cite{Jensen:2023yxy, Witten:2023qsv}, observers \cite{Harlow:2025pvj,Abdalla:2025gzn}, relational observables \cite{Goeller:2022rsx, Giddings:2025xym}, and observables defined using the canonical quantization approach to quantum gravity (for instance \cite{DeVuyst:2024uvd, DeVuyst:2024pop, Chandrasekaran:2022cip}) and the crossed product \cite{Witten:2023qsv,Jensen:2023yxy, Fewster:2024pur,AliAhmad:2024wja,Klinger:2023auu,Witten:2021unn}. 

A connection between the depth parameter in \cite{Gesteau:2024rpt} and the bulk-point singularity may suggest how string theory effects can enter. An intriguing possibility is if more general forms of causality, for example defined in some background-independent way from observables rather than light cones \cite{Witten:2023xze}, can furnish a generalized notion of time that exists entirely without geometry.

\begin{acknowledgments}
I thank Leonardo Badurina, Daniel Carney, Simon Caron-Huot, Yanbei Chen, Clifford Cheung, Flaminia Giacomini, Temple He, Philipp Hoehn, Manthos Karydas, Joon-Hwi Kim, Marc Klinger, Per Kraus, Antony Speranza, Sander Vermeulen, Jeffrey Wack, and Jordan Wilson-Gerow for discussions and comments on the draft, and Elliott Gesteau for initial collaboration. I also thank participants of the ``Observables in Quantum Gravity: from Theory to Experiment'' conference, held in January 2025 at the Aspen Center for Physics, for discussions. This work is supported by the Heising-Simons Foundation ``Observational Signatures of Quantum Gravity'' collaboration grant 2021-2817; by the D.O.E., Office of High Energy Physics, under Award No. DE-SC0011632; and by the Walter Burke Institute for Theoretical Physics. This work was performed in part at Aspen Center for Physics, which is supported by National Science Foundation grant PHY-2210452.
\end{acknowledgments}

\pagebreak

\bibliographystyle{utphys}
\bibliography{references}

\providecommand{\href}[2]{#2}\begingroup\raggedright\begin{thebibliography}{10}

\bibitem{DeVuyst:2024pop}
J.~De~Vuyst, S.~Eccles, P.~A. Hoehn, and J.~Kirklin, ``{Gravitational entropy is observer-dependent},'' \href{https://arxiv.org/abs/2405.00114}{{\ttfamily arXiv:2405.00114 [hep-th]}}.

\bibitem{DeVuyst:2024uvd}
J.~De~Vuyst, S.~Eccles, P.~A. Hoehn, and J.~Kirklin, ``{Crossed products and quantum reference frames: on the observer-dependence of gravitational entropy},'' \href{https://arxiv.org/abs/2412.15502}{{\ttfamily arXiv:2412.15502 [hep-th]}}.

\bibitem{Chandrasekaran:2022cip}
V.~Chandrasekaran, R.~Longo, G.~Penington, and E.~Witten, ``{An algebra of observables for de Sitter space},'' \href{https://dx.doi.org/10.1007/JHEP02(2023)082}{{\em JHEP} {\bfseries 02} (2023) 082}, \href{https://arxiv.org/abs/2206.10780}{{\ttfamily arXiv:2206.10780 [hep-th]}}.

\bibitem{Giddings:2025xym}
S.~B. Giddings, ``{Quantum gravity observables: observation, algebras, and mathematical structure},'' \href{https://arxiv.org/abs/2505.22708}{{\ttfamily arXiv:2505.22708 [hep-th]}}.

\bibitem{Donoghue:1995cz}
J.~F. Donoghue, ``{Introduction to the effective field theory description of gravity},'' in {\em {Advanced School on Effective Theories}}.
\newblock 6, 1995.
\newblock \href{https://arxiv.org/abs/gr-qc/9512024}{{\ttfamily arXiv:gr-qc/9512024}}.

\bibitem{Burgess:2003jk}
C.~P. Burgess, ``{Quantum gravity in everyday life: General relativity as an effective field theory},'' \href{https://dx.doi.org/10.12942/lrr-2004-5}{{\em Living Rev. Rel.} {\bfseries 7} (2004) 5--56}, \href{https://arxiv.org/abs/gr-qc/0311082}{{\ttfamily arXiv:gr-qc/0311082}}.

\bibitem{Kanno:2020usf}
S.~Kanno, J.~Soda, and J.~Tokuda, ``{Noise and decoherence induced by gravitons},'' \href{https://dx.doi.org/10.1103/PhysRevD.103.044017}{{\em Phys. Rev. D} {\bfseries 103} no.~4, (2021) 044017}, \href{https://arxiv.org/abs/2007.09838}{{\ttfamily arXiv:2007.09838 [hep-th]}}.

\bibitem{Parikh:2020fhy}
M.~Parikh, F.~Wilczek, and G.~Zahariade, ``{Signatures of the quantization of gravity at gravitational wave detectors},'' \href{https://dx.doi.org/10.1103/PhysRevD.104.046021}{{\em Phys. Rev. D} {\bfseries 104} no.~4, (2021) 046021}, \href{https://arxiv.org/abs/2010.08208}{{\ttfamily arXiv:2010.08208 [hep-th]}}.

\bibitem{Parikh:2020kfh}
M.~Parikh, F.~Wilczek, and G.~Zahariade, ``{Quantum Mechanics of Gravitational Waves},'' \href{https://dx.doi.org/10.1103/PhysRevLett.127.081602}{{\em Phys. Rev. Lett.} {\bfseries 127} no.~8, (2021) 081602}, \href{https://arxiv.org/abs/2010.08205}{{\ttfamily arXiv:2010.08205 [hep-th]}}.

\bibitem{Parikh:2020nrd}
M.~Parikh, F.~Wilczek, and G.~Zahariade, ``{The Noise of Gravitons},'' \href{https://dx.doi.org/10.1142/S0218271820420018}{{\em Int. J. Mod. Phys. D} {\bfseries 29} no.~14, (2020) 2042001}, \href{https://arxiv.org/abs/2005.07211}{{\ttfamily arXiv:2005.07211 [hep-th]}}.

\bibitem{Zurek:2020ukz}
K.~M. Zurek, ``{On vacuum fluctuations in quantum gravity and interferometer arm fluctuations},'' \href{https://dx.doi.org/10.1016/j.physletb.2022.136910}{{\em Phys. Lett. B} {\bfseries 826} (2022) 136910}, \href{https://arxiv.org/abs/2012.05870}{{\ttfamily arXiv:2012.05870 [hep-th]}}.

\bibitem{Verlinde:2019xfb}
E.~P. Verlinde and K.~M. Zurek, ``{Observational signatures of quantum gravity in interferometers},'' \href{https://dx.doi.org/10.1016/j.physletb.2021.136663}{{\em Phys. Lett. B} {\bfseries 822} (2021) 136663}, \href{https://arxiv.org/abs/1902.08207}{{\ttfamily arXiv:1902.08207 [gr-qc]}}.

\bibitem{Carney:2024wnp}
D.~Carney, M.~Karydas, and A.~Sivaramakrishnan, ``{Response of interferometers to the vacuum of quantum gravity},'' \href{https://arxiv.org/abs/2409.03894}{{\ttfamily arXiv:2409.03894 [hep-th]}}.

\bibitem{aasi2015advanced}
J.~Aasi {\em et~al.}, ``Advanced ligo,'' {\em Classical and quantum gravity} {\bfseries 32} no.~7, (2015) 074001.

\bibitem{Vermeulen:2024vgl}
S.~M. Vermeulen {\em et~al.}, ``{Photon-Counting Interferometry to Detect Geontropic Space-Time Fluctuations with GQuEST},'' \href{https://dx.doi.org/10.1103/PhysRevX.15.011034}{{\em Phys. Rev. X} {\bfseries 15} no.~1, (2025) 011034}, \href{https://arxiv.org/abs/2404.07524}{{\ttfamily arXiv:2404.07524 [gr-qc]}}.

\bibitem{Misner1973}
C.~W. {Misner}, K.~S. {Thorne}, and J.~A. {Wheeler}, {\em {Gravitation}}.
\newblock 1973.

\bibitem{ligo-note}
``Ligo technical document ligo-t0900043,''
\newblock \url{https://dcc.ligo.org/public/0000/T0900043/011/LIGO-T0900043-11.pdf}.

\bibitem{Maggiore:2007ulw}
M.~Maggiore, \href{https://dx.doi.org/10.1093/acprof:oso/9780198570745.001.0001}{{\em {Gravitational Waves. Vol. 1: Theory and Experiments}}}.
\newblock Oxford University Press, 2007.

\bibitem{Donnelly:2015hta}
W.~Donnelly and S.~B. Giddings, ``{Diffeomorphism-invariant observables and their nonlocal algebra},'' \href{https://dx.doi.org/10.1103/PhysRevD.93.024030}{{\em Phys. Rev. D} {\bfseries 93} no.~2, (2016) 024030}, \href{https://arxiv.org/abs/1507.07921}{{\ttfamily arXiv:1507.07921 [hep-th]}}. [Erratum: Phys.Rev.D 94, 029903 (2016)].

\bibitem{Donnelly:2016rvo}
W.~Donnelly and S.~B. Giddings, ``{Observables, gravitational dressing, and obstructions to locality and subsystems},'' \href{https://dx.doi.org/10.1103/PhysRevD.94.104038}{{\em Phys. Rev. D} {\bfseries 94} no.~10, (2016) 104038}, \href{https://arxiv.org/abs/1607.01025}{{\ttfamily arXiv:1607.01025 [hep-th]}}.

\bibitem{Goeller:2022rsx}
C.~Goeller, P.~A. Hoehn, and J.~Kirklin, ``{Diffeomorphism-invariant observables and dynamical frames in gravity: reconciling bulk locality with general covariance},'' \href{https://arxiv.org/abs/2206.01193}{{\ttfamily arXiv:2206.01193 [hep-th]}}.

\bibitem{Perlick1987}
V.~Perlick, ``Characterization of standard clocks by means of light rays and freely falling particles,'' \href{https://dx.doi.org/10.1007/BF00759142}{{\em General Relativity and Gravitation} {\bfseries 19} no.~11, (Nov., 1987) 1059--1073}. \url{https://doi.org/10.1007/BF00759142}.

\bibitem{Perlick:2007np}
V.~Perlick, ``{On the Radar method in general-relativistic spacetimes},'' \href{https://dx.doi.org/10.1007/978-3-540-34377-6_5}{{\em Astrophys. Space Sci. Libr.} {\bfseries 349} (2008) 131--152}, \href{https://arxiv.org/abs/0708.0170}{{\ttfamily arXiv:0708.0170 [gr-qc]}}.

\bibitem{maggiore2008gravitational}
M.~Maggiore, {\em Gravitational waves}, vol.~2.
\newblock Oxford university press, 2008.

\bibitem{Takesaki:1970aki}
M.~Takesaki, \href{https://dx.doi.org/10.1007/bfb0065832}{{\em Tomita's Theory of Modular Hilbert Algebras and Its Applications}}.
\newblock Lecture Notes in Mathematics. Springer-Verlag, 1970.

\bibitem{Lee:2024oxo}
V.~S.~H. Lee and K.~M. Zurek, ``{Proper Time Observables of General Gravitational Perturbations in Laser Interferometry-based Gravitational Wave Detectors},'' \href{https://arxiv.org/abs/2408.03363}{{\ttfamily arXiv:2408.03363 [hep-ph]}}.

\bibitem{Sivaramakrishnan:2024ydy}
A.~Sivaramakrishnan, ``{Correlators of Worldline Proper Length},'' \href{https://arxiv.org/abs/2406.17205}{{\ttfamily arXiv:2406.17205 [hep-th]}}.

\bibitem{Ooguri:1985nv}
H.~Ooguri, ``{Spectrum of Hawking Radiation and Huygens' Principle},'' \href{https://dx.doi.org/10.1103/PhysRevD.33.3573}{{\em Phys. Rev. D} {\bfseries 33} (1986) 3573}.

\bibitem{Meltzer:2020qbr}
D.~Meltzer and A.~Sivaramakrishnan, ``{CFT unitarity and the AdS Cutkosky rules},'' \href{https://dx.doi.org/10.1007/JHEP11(2020)073}{{\em JHEP} {\bfseries 11} (2020) 073}, \href{https://arxiv.org/abs/2008.11730}{{\ttfamily arXiv:2008.11730 [hep-th]}}.

\bibitem{Cheung:2022pdk}
C.~Cheung, J.~Parra-Martinez, and A.~Sivaramakrishnan, ``{On-shell correlators and color-kinematics duality in curved symmetric spacetimes},'' \href{https://dx.doi.org/10.1007/JHEP05(2022)027}{{\em JHEP} {\bfseries 05} (2022) 027}, \href{https://arxiv.org/abs/2201.05147}{{\ttfamily arXiv:2201.05147 [hep-th]}}.

\bibitem{Aalsma:2025bcg}
L.~Aalsma and S.-E. Bak, ``{Modular Fluctuations in Cosmology},'' \href{https://arxiv.org/abs/2503.04886}{{\ttfamily arXiv:2503.04886 [hep-th]}}.

\bibitem{Jensen:2023yxy}
K.~Jensen, J.~Sorce, and A.~J. Speranza, ``{Generalized entropy for general subregions in quantum gravity},'' \href{https://dx.doi.org/10.1007/JHEP12(2023)020}{{\em JHEP} {\bfseries 12} (2023) 020}, \href{https://arxiv.org/abs/2306.01837}{{\ttfamily arXiv:2306.01837 [hep-th]}}.

\bibitem{Witten:2023qsv}
E.~Witten, ``{Algebras, regions, and observers},'' {\em Proc. Symp. Pure Math.} {\bfseries 107} (2024) 247--276, \href{https://arxiv.org/abs/2303.02837}{{\ttfamily arXiv:2303.02837 [hep-th]}}.

\bibitem{Harlow:2025pvj}
D.~Harlow, M.~Usatyuk, and Y.~Zhao, ``{Quantum mechanics and observers for gravity in a closed universe},'' \href{https://arxiv.org/abs/2501.02359}{{\ttfamily arXiv:2501.02359 [hep-th]}}.

\bibitem{Abdalla:2025gzn}
A.~I. Abdalla, S.~Antonini, L.~V. Iliesiu, and A.~Levine, ``{The gravitational path integral from an observer's point of view},'' \href{https://arxiv.org/abs/2501.02632}{{\ttfamily arXiv:2501.02632 [hep-th]}}.

\bibitem{Fewster:2024pur}
J.~C. Fewster, D.~W. Janssen, L.~D. Loveridge, K.~Rejzner, and J.~Waldron, ``{Quantum Reference Frames, Measurement Schemes and the Type of Local Algebras in Quantum Field Theory},'' \href{https://dx.doi.org/10.1007/s00220-024-05180-7}{{\em Commun. Math. Phys.} {\bfseries 406} no.~1, (2025) 19}, \href{https://arxiv.org/abs/2403.11973}{{\ttfamily arXiv:2403.11973 [math-ph]}}.

\bibitem{AliAhmad:2024wja}
S.~Ali~Ahmad, W.~Chemissany, M.~S. Klinger, and R.~G. Leigh, ``{Quantum reference frames from top-down crossed products},'' \href{https://dx.doi.org/10.1103/PhysRevD.110.065003}{{\em Phys. Rev. D} {\bfseries 110} no.~6, (2024) 065003}, \href{https://arxiv.org/abs/2405.13884}{{\ttfamily arXiv:2405.13884 [hep-th]}}.

\bibitem{Klinger:2023auu}
M.~S. Klinger and R.~G. Leigh, ``{Crossed products, conditional expectations and constraint quantization},'' \href{https://dx.doi.org/10.1016/j.nuclphysb.2024.116622}{{\em Nucl. Phys. B} {\bfseries 1006} (2024) 116622}, \href{https://arxiv.org/abs/2312.16678}{{\ttfamily arXiv:2312.16678 [hep-th]}}.

\bibitem{Witten:2021unn}
E.~Witten, ``{Gravity and the crossed product},'' \href{https://dx.doi.org/10.1007/JHEP10(2022)008}{{\em JHEP} {\bfseries 10} (2022) 008}, \href{https://arxiv.org/abs/2112.12828}{{\ttfamily arXiv:2112.12828 [hep-th]}}.

\bibitem{Gesteau:2024rpt}
E.~Gesteau and H.~Liu, ``{Toward stringy horizons},'' \href{https://arxiv.org/abs/2408.12642}{{\ttfamily arXiv:2408.12642 [hep-th]}}.

\bibitem{Witten:2023xze}
E.~Witten, ``{A background-independent algebra in quantum gravity},'' \href{https://dx.doi.org/10.1007/JHEP03(2024)077}{{\em JHEP} {\bfseries 03} (2024) 077}, \href{https://arxiv.org/abs/2308.03663}{{\ttfamily arXiv:2308.03663 [hep-th]}}.

\bibitem{Tinsley2016}
J.~N. Tinsley, M.~I. Molodtsov, R.~Prevedel, D.~Wartmann, J.~Espigul{\'e}-Pons, M.~Lauwers, and A.~Vaziri, ``Direct detection of a single photon by humans,'' {\em Nature Communications} {\bfseries 7} no.~1, (2016) 12172.

\bibitem{Herrmann:2024yai}
E.~Herrmann, M.~Kologlu, and I.~Moult, ``{Energy Correlators in Perturbative Quantum Gravity},'' \href{https://arxiv.org/abs/2412.05384}{{\ttfamily arXiv:2412.05384 [hep-th]}}.

\bibitem{Moult:2025nhu}
I.~Moult and H.~X. Zhu, ``{Energy Correlators: A Journey From Theory to Experiment},'' \href{https://arxiv.org/abs/2506.09119}{{\ttfamily arXiv:2506.09119 [hep-ph]}}.

\bibitem{Gary:2009ae}
M.~Gary, S.~B. Giddings, and J.~Penedones, ``{Local bulk S-matrix elements and CFT singularities},'' \href{https://dx.doi.org/10.1103/PhysRevD.80.085005}{{\em Phys. Rev. D} {\bfseries 80} (2009) 085005}, \href{https://arxiv.org/abs/0903.4437}{{\ttfamily arXiv:0903.4437 [hep-th]}}.

\bibitem{Polchinski:1999yd}
J.~Polchinski, L.~Susskind, and N.~Toumbas, ``{Negative energy, superluminosity and holography},'' \href{https://dx.doi.org/10.1103/PhysRevD.60.084006}{{\em Phys. Rev. D} {\bfseries 60} (1999) 084006}, \href{https://arxiv.org/abs/hep-th/9903228}{{\ttfamily arXiv:hep-th/9903228}}.

\bibitem{Heemskerk:2009pn}
I.~Heemskerk, J.~Penedones, J.~Polchinski, and J.~Sully, ``{Holography from Conformal Field Theory},'' \href{https://dx.doi.org/10.1088/1126-6708/2009/10/079}{{\em JHEP} {\bfseries 10} (2009) 079}, \href{https://arxiv.org/abs/0907.0151}{{\ttfamily arXiv:0907.0151 [hep-th]}}.

\bibitem{Penedones:2010ue}
J.~Penedones, ``{Writing CFT correlation functions as AdS scattering amplitudes},'' \href{https://dx.doi.org/10.1007/JHEP03(2011)025}{{\em JHEP} {\bfseries 03} (2011) 025}, \href{https://arxiv.org/abs/1011.1485}{{\ttfamily arXiv:1011.1485 [hep-th]}}.

\bibitem{Okuda:2010ym}
T.~Okuda and J.~Penedones, ``{String scattering in flat space and a scaling limit of Yang-Mills correlators},'' \href{https://dx.doi.org/10.1103/PhysRevD.83.086001}{{\em Phys. Rev. D} {\bfseries 83} (2011) 086001}, \href{https://arxiv.org/abs/1002.2641}{{\ttfamily arXiv:1002.2641 [hep-th]}}.

\bibitem{Maldacena:2015iua}
J.~Maldacena, D.~Simmons-Duffin, and A.~Zhiboedov, ``{Looking for a bulk point},'' \href{https://dx.doi.org/10.1007/JHEP01(2017)013}{{\em JHEP} {\bfseries 01} (2017) 013}, \href{https://arxiv.org/abs/1509.03612}{{\ttfamily arXiv:1509.03612 [hep-th]}}.

\bibitem{Caron-Huot:2022lff}
S.~Caron-Huot, ``{Holographic cameras: an eye for the bulk},'' \href{https://dx.doi.org/10.1007/JHEP03(2023)047}{{\em JHEP} {\bfseries 03} (2023) 047}, \href{https://arxiv.org/abs/2211.11791}{{\ttfamily arXiv:2211.11791 [hep-th]}}.

\bibitem{Caron-Huot:2025hmk}
S.~Caron-Huot, J.~Chakravarty, and K.~Namjou, ``{Looking at bulk points in general geometries},'' \href{https://arxiv.org/abs/2502.14963}{{\ttfamily arXiv:2502.14963 [hep-th]}}.

\bibitem{Engelhardt:2016wgb}
N.~Engelhardt and G.~T. Horowitz, ``{Towards a Reconstruction of General Bulk Metrics},'' \href{https://dx.doi.org/10.1088/1361-6382/34/1/015004}{{\em Class. Quant. Grav.} {\bfseries 34} no.~1, (2017) 015004}, \href{https://arxiv.org/abs/1605.01070}{{\ttfamily arXiv:1605.01070 [hep-th]}}.

\end{thebibliography}\endgroup

%\clearpage

\clearpage 
\pagebreak

\appendix

\section{Proper time from detector correlators}
\label{detectors}

Here we sketch how proper time can be defined operationally in a more realistic setting, that is, via measurements made using detectors. Suppose observers can measure the value of a scalar quantum field $\phi$ and its correlation functions  $\braket{\phi(x_1) \cdots \phi(x_n)}$. This models  measurement in actual experiments.\footnote{In fact, single-photon detectors may model what quantum gravity effects would actually look like if they were large enough to be detected in human vision \cite{Tinsley2016}.} The detector in the laser interferometer GQuEST, for example, is a superconducting nanowire single-photon detector \cite{Vermeulen:2024vgl}. Effectively, the detector measures power deposited by the electromagnetic field as a function of time. In our case, the electromagnetic field is modelled by $\phi$ and the light sources are modelled by pointlike excitations of $\phi$. See also \cite{Herrmann:2024yai, Moult:2025nhu}. 

Correlation functions in quantum field theory even in curved space are expected to have singularities in Lorentzian signature only when points are null separated. Light-cone singularities in the various correlators (e.g. time-ordered, Wightman, retarded, advanced) capture a precise notion of causality, that signals in local theories cannot propagate outside light cones.

We now explain how $\tau_3-\tau_1$ defined by a lightbulb clock is extracted from detector correlators. The connected vacuum  correlator $\braket{\phi(x_1(\tau_1) )\phi(x_1(\tau))}_c$ involves operators inserted at timelike separated points and so it has no light-cone singularity. However, in the lightbulb clock protocol, a $\phi$ excitation starts at $x_1(\tau_1)$ and through the actions of the second observer, the original excitation induces a second $\phi$ excitation to arrive at $x_1(\tau)$ for some $\tau$. This implies $\braket{\phi(x_1(\tau_1) )\phi(x_1(\tau))}_c$ has a light-cone singularity. The location $\tau$ of this worldline bulk-point singularity, as we will call it, is defined by
\begin{equation}
\tau~~\text{such that}~~\braket{\phi(x_1(\tau_1)) \phi(x_1(\tau))}_c = \infty
\label{bulkpointcondition}
\end{equation}
in the presence of the second observer. This condition fixes $\tau = \tau_3$ as defined in Section \ref{lightbulbclock}. Because the bulk point condition and therefore $\tau_3$ is determined by the detector output,\footnote{This is sometimes called a field relational observable. See \cite{Giddings:2025xym} for a recent review.} it is manifestly diffeomorphism invariant.\footnote{Finite resolution in time can be included by using time-averaged operators, for which the bulk-point singularity becomes an approximate notion.} 

\begin{figure}[ht]
    \centering
    \includegraphics[width=0.5 \textwidth]{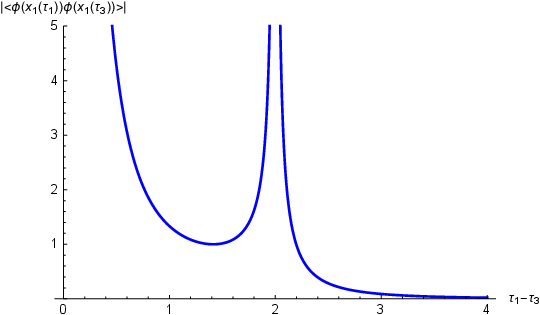}
    \caption{Graph of detector output, specifically $|\braket{\phi(x_1(\tau_1))\phi(x_1(\tau_3))}|$ from \eqref{twopointfunction_mirror} with $L=1$. The singularity at $\tau_3-\tau_1 = 2L$ corresponds to reflection from the mirror. This shows how the time of arrival $\tau_3$ can be read off from what detectors measure.}
    \label{detector_plot}
\end{figure}

Detector correlators on different worldlines fix $\tau_2$ via a similar condition. There exists a $\tau$ such that $\braket{\phi(x_1(\tau_1)) \phi(x_2(\tau))}_c = \infty$, and this condition fixes $\tau = \tau_2$. We conclude that detector correlators can therefore implement the lightbulb clock protocol. See figure \ref{detector_plot} for a graph of such a correlator and Appendix \ref{spectraldata} for an explicit example in flat space.

Note that worldline bulk-point singularities of detector correlators exist only if the bulk theory is local. This criterion for bulk locality has been studied in holography, where even nonperturbative quantum gravity effects can be captured \cite{Gary:2009ae, Polchinski:1999yd, Heemskerk:2009pn,Penedones:2010ue, Okuda:2010ym,Maldacena:2015iua}. See also the camera protocol in \cite{Caron-Huot:2022lff, Caron-Huot:2025hmk} and the light cone cuts in \cite{Engelhardt:2016wgb} for progress along similar lines in holography.

\section{Singularities in detector correlators from spectral data}
\label{spectraldata}

In experimental applications, detector correlations are often analysed in spectral or frequency space, which is connected to the worldline bulk-point singularity condition. Suppose we have a planar mirror located at $z=L$. Using the method of images, the two-point Wightman function of a free scalar field obeys $\braket{\phi(x_1) \phi(x_2)}_{m} = 0$ when $x_i^\mu = (t_i,0,0,L)$, which implies
\begin{equation}
\braket{\phi(x_1) \phi(x_2)}_{m} =G(x_1,x_2)-G(x_1,x_2'),
\label{twopointfunction_mirror}
\end{equation}
with $z_2' = 2L-z_2$.  We now choose $x_1,x_2$ to be on the same zero-velocity geodesic, $\mathbf{x}_1=\mathbf{x}_2$, with $x^\mu = (t,\mathbf{x})$. The worldline bulk-point singularity comes from the second term in \eqref{twopointfunction_mirror}, which is precisely what encodes the presence of the mirror.

The vacuum Wightman function is
\begin{equation}
G(x_1,x_2) = \frac{1}{|\mathbf{x}_{12}|}\int_0^\infty dk \sin(k |\mathbf{x}_{12}|)e^{-i kt_{12}},
\end{equation}
or
\begin{equation}
\braket{\phi(x_1) \phi(x_2)}_{m} = -\int_0^\infty dk \sin(2 k L)e^{-i k t_{12}}.
\end{equation}
The elapsed proper time in this setup, $2L$, can be read off from the spectral function, $-\sin(2 k L)$.

\section{Perturbed photon trajectory}
\label{propertime}
Here we solve for $x_p$ perturbatively. We begin by writing $ds^2=0$ as
\begin{equation}
v^2=1- h^{ij}(x_p) v_i v_j.
\label{nullvelocityEOM}
\end{equation}
Expanding \eqref{nullvelocityEOM} to second order,
\begin{align}
&1  +  2 v^{(0)} (v^{(1)}+v^{(2)})+(v^{(1)})^2
\nonumber
\\
&~~~=1- h^{ij} ( v^{(0)}_i v^{(0)}_j + 2 v^{(0)}_i v^{(1)}_j) + x_p^{(1),\mu}\partial_\mu h^{ij}  v^{(0)}_i v^{(0)}_j,
\end{align}
where we used that $h^{ij}=h^{ji}$ and for brevity, we denote $h^{ij} = h^{ij}(x^{(0)}_p)$. We also use that $v_{i}^{(0)} = (0,0,\pm 1)$, where outgoing is $(+)$ and incoming is $(-)$, and $(v_i^{(0)})^2 = 1$. At first order,
\begin{equation}
\mathcal{O}(h):~~~ v^{(0)} v^{(1)}
=- \frac{1}{2} h^{ij}  v^{(0)}_i v^{(0)}_j,
\end{equation}  
which implies $\pm v^{(1)}_z = -\frac{1}{2}h^{zz}$. 
At second order,
\begin{align}
\mathcal{O}(h^2): ~~ &v^{(0)} v^{(2)}+\frac{1}{2}(v^{(1)})^2 
\nonumber
\\
&~~=-  h^{ij} v^{(0)}_i v^{(1)}_j - \frac{1}{2}x_p^{(1),\mu}\partial_\mu h^{ij}  v^{(0)}_i v^{(0)}_j,
\end{align}
or
\begin{equation}
\mathcal{O}(h^2): ~~ \pm v^{(2)}_z+\frac{1}{2}(v^{(1)})^2 
=\mp h^{z j}  v^{(1)}_j - \frac{1}{2}x_p^{(1),\mu}\partial_\mu h^{zz}.
\end{equation}
This involves components of $v^{(1)}, x^{(1)}$ we have not yet solved for. The deflection of the null ray transverse to the $t-z$ plane at first order did not affect elapsed times at that order, but can at second order. The required components can be obtained by solving the geodesic equation,
\begin{equation}
\mathcal{O}(h): ~~  \dot{v}^{(1),i} + \frac{1}{2}\eta^{im}(2(\partial_z+v^{(0)}_z\partial_t) h_{mz} -\partial_m h_{zz}) = 0,
\end{equation}
whose solution is
\begin{equation}
~~v^{(1)}_i  = -\frac{1}{2}\int_0^{2L}  (2(\partial_z+v^{(0)}_z\partial_t) h_{iz}-\partial_i h_{zz})dt.
\end{equation}
Integrating a second time computes $x^{(1)}_p$. This is subject to the boundary conditions that ensure the null rays begin and end at the same location in the transverse plane, therefore reaching the two timelike geodesics, 
\begin{equation}
0=\int_0^L v^{(1)}_i dt = \int_L^{2L} v^{(1)}_i dt
\end{equation}
for all $i$ corresponding to the transverse directions. 

We will not need the explicit form of the transverse components of $v^{(1)}$ for our purposes, only that all $v^{(1)}_i$ are linear in $h$, which implies $v^{(2)}$ is composed of quadratic functions of $h$ and its time integrals.

\section{Additional review from \cite{Carney:2024wnp}}
\label{propertimeoperator}

To derive the proper time Wightman function in transverse-traceless gauge, we use
\begin{equation}
h_{\mu\nu} = \ell_p \sum_s \int \frac{d^3 \mathbf{k}}{(2 \pi)^3 \sqrt{2 E_k}} (\epsilon_{\mu\nu}^s(\mathbf{k}) e^{i k x} a_{\mathbf{k},s} + c.c.),
\end{equation}
with two polarizations $s=1,2$, the commutator $[a_{\mathbf{k},s},a^\dagger_{\mathbf{k}',s'} ] = \delta^{(3)}(\mathbf{k}-\mathbf{k}') \delta_{s s'}$, and the normalization $\ell_p^2 = 32 \pi G_N $. The momentum is on shell, $k^0 = |\mathbf{k}|$. As the unperturbed setup (or interferometer baseline) is entirely in the $z$ plane, we need only the polarization sum $\sum_s |\epsilon^s_{zz}|^2 = \sin^4 \theta$. This leads to
\begin{equation}
\braket{h_{zz}(x)h_{zz}(x')} = 
\frac{\ell^2_p}{2(2 \pi)^3} \int \frac{d^3 \mathbf{k}}{|\mathbf{k}|} \sin^4 \theta  e^{i k (x-x')} ,
\end{equation}
or moving to spherical coordinates and changing notation to $|\mathbf{k}| \rightarrow k$,
\begin{equation}
\braket{h_{zz}(x)h_{zz}(x')} = 
\frac{\ell_p^2}{2(2 \pi)^3} \int d\phi \,d\theta \int_0^\infty dk~ k \sin^5 \theta  e^{ i p  (x-x')} ,
\end{equation}
with $p = k(1, \sin \theta \cos \phi, \sin \theta \sin \phi, \cos \theta)$. As explained in Appendix A of \cite{Carney:2024wnp}, the Wightman function is defined by first taking $t \rightarrow t-i\epsilon$, which regulates the large-$k$ behavior of the integral, then evaluating the integral, and finally taking the $\epsilon \rightarrow 0^+$ limit.

\section{Anticommutator of elapsed time}
\label{anticommutator}
For elapsed time $\tau$, the closed-form expression for $2\,\text{Re}\braket{\tau(t)\tau(0)}_c$ is
\begin{align}
&2\,\text{Re}\braket{\tau(t)\tau(0)}_c
= \frac{8 G_N}{9 \pi}
\bigg(
-18 (1 + 6 u^2) \log(2 L u)
\nonumber
\\
&+(9 - 36 u + 54 u^2 - 24 u^3) \log(2 L |u-1|)-34 - 48 u^2
\nonumber
\\
&+(9 + 36 u + 54 u^2 + 24 u^3)\log(2 L (1 + u))
\bigg)
\end{align}
for all $u>0$. For $u \gg 1$, we have $2\,\text{Re}\braket{\tau(t)\tau(0)}_c \approx -\frac{32 G_N}{ 15 \pi u^2}$. As expected, this is the same large-time decay as the vacuum two-point function of $h$. The singularities at $u=0,1$ are usually regulated by accounting for finite experimental resolution, as explained in \cite{Carney:2024wnp}.

\end{document}